\documentclass[english,pra,preprint,superscriptaddress]{revtex4-1}
\usepackage{amsmath,amssymb,bm}
\usepackage{graphicx}
\usepackage{dcolumn}
\usepackage{ifthen}
\usepackage{babel}
\usepackage{color}
\usepackage{hyperref}
\hypersetup{linktoc=page}
\hypersetup{
    colorlinks,
    citecolor=blue,    filecolor=blue,
    linkcolor=blue,     urlcolor=blue
}

\begin{document}
\title{Theory of femtosecond strong field ion excitation and subsequent lasing in N$_2^+$}
\author{Vladimir T. Tikhonchuk}
\affiliation{University of Bordeaux-CNRS-CEA, CELIA, UMR 5107, 33405 Talence, France}
\email{tikhonchuk@u-bordeaux.fr}
\affiliation{ELI-Beamlines, Institute of Physics, Czech Academy of Sciences,  25241 Doln\'{i} B\v{r}e\v{z}any, Czech Republic}
\author{Yi Liu}
\affiliation{Shanghai Key Lab of Modern Optical System, University of Shanghai for Science and Technology, 200093 Shanghai, China}
\author{Rostyslav Danylo}
\affiliation{Shanghai Key Lab of Modern Optical System, University of Shanghai for Science and Technology, 200093 Shanghai, China}
\affiliation{Laboratoire d'Optique Appliqu\'ee, ENSTA ParisTech, CNRS, Ecole Polytechnique, Institut Polytechnique de Paris, 91762 Palaiseau, France}
\email{rostyslav.danylo@ensta-paristech.fr}
\author{Aur\'{e}lien Houard}
\affiliation{Laboratoire d'Optique Appliqu\'ee, ENSTA ParisTech, CNRS, Ecole Polytechnique, Institut Polytechnique de Paris, 91762 Palaiseau, France}
\author{Andr\'e Mysyrowicz}
\affiliation{Laboratoire d'Optique Appliqu\'ee, ENSTA ParisTech, CNRS, Ecole Polytechnique, Institut Polytechnique de Paris, 91762 Palaiseau, France}
\date{\today}

\begin{abstract}
Delayed cavity-free forward lasing at the wavelengths of 391 and 428~nm was observed in recent experiments in air or pure nitrogen pumped with an intense femtosecond laser pulse at wavelength of 800~nm. The mechanism responsible for the lasing is highly controversial. In this article we explain the delayed emission by the presence of long-lived polarizations coupling simultaneously ground state X$^2\Sigma_g^+$ to states A$^2\Pi_u$ and B$^2\Sigma_u^+$ of singly ionized nitrogen molecules N$_2^+$. Ionization of neutral nitrogen molecules in a strong laser field and subsequent ion excitation are described by a system of Bloch equations providing a distribution of ions in the ground and excited states A and B at the end of the laser pulse. The delayed signal amplification at the B-X transition wavelength is described by a system of Maxwell-Bloch equations with polarization coupling maintained by a weak laser post-pulse. Two regimes of signal amplification are identified: a signal of a few ps duration at low gas pressures and a short (sub-picosecond) signal at high gas pressures. The theoretical model compares favorably with experimental results.
\end{abstract}

\maketitle

\section{Introduction: lasing without inversion}\label{sec1}
Several experiments on the interaction of a strong ultra-short laser pulse at 800~nm with molecular nitrogen~\cite{Yao2013, Liu2013, Li2014, Point2014, Xu2015, Yao2016, Zhang2017} report on a robust cavity-free lasing in the forward direction at the wavelengths 391 or 428~nm, corresponding to transitions from the excited state B$^2\Sigma_u^+$ to the ground state X$^2\Sigma_g^+$ of N$_2^+$ with vibrational level 0 or 1. It was observed that a femtosecond seed signal at 391 or 428~nm, injected a few ps after the pump pulse, is amplified by two-three orders of magnitude. Furthermore, the lasing emission is delayed from the seed pulse by a few ps. Several explanations for this lasing have been proposed so far. Fast population inversion between B and X due to depletion of X by the pump pulse inducing a transfer of population from X to the intermediate third level  A$^2\Pi_u$ has been discussed in Refs.~\cite{Xu2015, Yao2016}. However, these authors did not offer an explanation for the retarded emission, which is much longer that the pump pulse duration. Increase of the population at the upper level by multiple electron recollisions was suggested in~\cite{Liu15}. However, it was shown later that this process itself cannot produce an optical gain needed for lasing~\cite{Tikhonchuk17}.

It is known that lasing without population inversion is possible in a V-scheme, which involves a third level resonantly driven by the pump and coupled to the excited state by a quantum interference~\cite{Kocharovskaya1988, Harris1988, Scully1989}. Such a V-scheme, studied in Refs.~\cite{Mompart1998, Malyshev1998, Kozlov1999} and more recently in Ref.~\cite{Scully2013}, can be applied to the interpretation of lasing of nitrogen molecular ions N$_2^+$ driven by a ultrashort laser pulse with peak intensity in the range of a few $10^{14}$~W/cm$^2$, but one needs to explain how the third level is driven and what are the conditions for obtaining optical amplification. Recently we have developed a theoretical model that is capable of describing the delayed lasing~\cite{Liu19}. The theoretical results are in agreement with experimental data about the temporal shape of the amplified signal and the gas pressure dependence of the gain. The present paper explains in more details this theory: The signal amplification in the B-X transition of nitrogen molecular ions is described by a two-step process: (i) the interaction of a short and intense pump laser pulse at the wavelength 800~nm with a nitrogen gas leads to partial ionization of the nitrogen molecules and partial excitation of the molecular ions to the upper states A and B; (ii) the coherent cross-coupling between excited ion states A and B opens the possibility of retarded signal amplification. The following condition must be satisfied in order to obtain gain: population at level B must be larger than at level A but smaller than at level X; the coherent cross-coupling between A and B must be maintained by a weak 800~nm post-pulse of a few ps duration.

The paper is organized as follows. Section~\ref{sec2} addresses the problem of ionization of nitrogen molecules and excitation of the ions by the main laser pulse. While direct ionization into excited ionic states has a low probability, population transfer to these excited states can be quite efficient if the transition frequencies are of the same order of magnitude as the corresponding Rabi frequencies. The calculations show that in the range of intensities around $10^{14}$~W/cm$^2$ population at level B is always lower than that at level X(0). Nevertheless, as we show in Sec.~\ref{sec3}, signal amplification after the end of the laser pulse is possible provided that the main laser pulse is followed by a weak coherent post-pulse of a few ps duration, which maintains the  A-X polarization. The presence of a post-pulse is consistent with experimental observations~\cite{Liu19}. The temporal evolution of populations in these three resonantly coupled levels and the evolution of electromagnetic fields is described by a system of Maxwell-Bloch equations enveloped over the transition frequencies. This long-lived mutual coherence makes the system unstable: it may generate an emission corresponding to the B-X transition or amplify a seed injected at the corresponding wavelength in the absence of population inversion between B and X. Depending on the post-pulse fluence and gas pressure, this amplification may proceed in two regimes: either parametric signal amplification at low pressures or soliton formation at high pressures. Section~\ref{sec4} presents an analysis of numerical simulations with a particular emphasis on the dependence of the amplification process on gas pressure and laser post-pulse amplitude. Section~\ref{sec5} compares our theoretical results with experiments.  Section~\ref{sec6} presents our conclusions.

\section{Ion excitation in a strong laser field}\label{sec2}
Here we consider the interaction of the main laser pulse at 800~nm with a homogeneous nitrogen gas. The laser pulse intensity, on the order of $10^{14}$~W/cm$^2$, is sufficiently strong to create a plasma filament, and it is assumed that it is not appreciably modified by the gas ionization and ion excitation. Therefore, we consider the interaction of a given laser field with a single nitrogen molecule. The physical processes that we are interested in are: (i) ionization of the neutral nitrogen molecule from the neutral ground state to the ionic ground X and excited A and B states and (ii) subsequent transitions during the laser pulse between the ground and excited ionic states.

\begin{figure}[!ht]
\setlength{\unitlength}{1mm}
\centerline{\begin{picture}(55,30)
\put(0,0){\line(1,0){20}}
\put(22,-1){0}
\put(1,2){\line(1,0){20}}
\put(24,1){1}
\put(-10,2){X$^2\Sigma_g^+$}
\put(-10,30){\line(1,0){15}}
\put(-15,25){B$^2\Sigma_u^+$}
\put(7,29){0}
\put(13,19){\line(1,0){20}}
\put(12,17){\line(1,0){20}}
\put(11,15){\line(1,0){20}}
\put(10,13){\line(1,0){20}}
\put(16,22){A$^2\Pi_u$}
\put(35,18){3}
\put(34,16){2}
\put(33,14){1}
\put(32,12){0}
\thicklines
\put(8,2){\vector(-1,3){9.3}}
\put(12,2){\vector(1,2){8.5}}
\put(6,0){\vector(-1,3){10}}
\put(13,0){\vector(1,2){8.5}}
\put(-10,15){$\omega_{bx0}$}
\put(18,5){$\omega_{ax0}$}
\put(4,18){$\omega_{bx1}$}
\put(7,9){$\omega_{ax1}$}
\thinlines
\multiput(14.5,16.5)(-2.4,2.4){6}{\vector(-1,1){1.57}}
\multiput(15,19)(-2.4,2.4){5}{\vector(-1,1){1.5}}
\end{picture}}
\caption{Scheme of the electronic levels in the molecule N$_2^+$. Numbers near the lines indicate the vibrational mode, rotational splitting is neglected. Energy gap for transition B(0)-X(0) is 3.20~eV, energy gap for  transition A(2)-X(0) is 1.58~eV, energy splitting between vibrational X levels is 0.27~eV, energy splitting between the vibrational A levels 2 and 3 is 0.25~eV.  Resonant couplings to the laser corresponding to transitions X(0)-A(2)-B(0) and X(1)-A(3)-B(0) are shown with thick arrows. Polarization cross-coupling between A and B levels is shown with dashed arrows. } \label{fig1}
\end{figure}
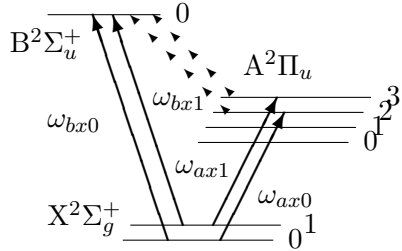

Fig.~\ref{fig1} shows the V-scheme of energy levels of N$_2^+$ containing the ground levels X$^2\Sigma_g^+$(0,1), the excited level B$^2\Sigma_u^+$(0) and a series of A$^2\Pi_u(v)$ levels with different vibrational quantum numbers varying from $v = 0$ to 3. The resonant couplings considered in what follows correspond to transitions X(0)-A(2) and X(1)-A(3).

Ionization of the nitrogen molecules is described by the PPT model~\cite{PPT1966, Popov2004} with the ionization probability $w_{\rm PPT} (U,E)g(\theta_n)$ depending on the ionization potential $U$, instantaneous laser electric field $E(t)$, and angle $\theta_n$ between its direction and molecule axis. The angular dependence is interpolated according to~\cite{Petretti2010} by a function $g(\theta_n)= 0.45+ 0.95\,\cos^2\theta_n+ 1.17\,\cos^4\theta_n$, which decreases from the maximum value 2.57 for the parallel orientation, $\theta=0^\circ$, to 0.45 for the perpendicular orientation, $\theta=90^\circ$, with an average value of 1. The ionization energy of level X(0) $U_{X0}=15.57$~eV is comparable to that of level X(1), $U_{X1}=15.84$~eV, so that both levels have to be considered. As the probability of tunnel ionization decreases exponentially with the transition energy, direct ionization into excited states A and B, separated by the energy gap of 1.5 and 3~eV, respectively, is small, but we account for it for completeness.  

The dipolar moments $\mu_{ax}\simeq 0.25$~at.u. and $\mu_{bx}\simeq 0.75$~at.u. of A-X and B-X transitions~\cite{Langhoff1987, Langhoff1988} correspond to the coupling energy $\mu E$ on the order of a few eV for a laser intensity of $10^{14}$~W/cm$^2$. This value is comparable to the energies of transitions. Therefore, several levels can be excited simultaneously irrespectively of the resonance conditions. Non-resonant excitation of two- and three-level systems in a strong laser field was considered in Refs.~\cite{Xu2015, Yao2016, Zhang2017}. Here we extend this approach by considering excitation in three states: B(0), A(2) and A(3). Triplet X(0)-A(2)-B(0) with transition energies 1.58 and 3.20~eV, corresponds to the lasing signal observed at 391~nm, another triplet X(1)-A(3)-B(0) with transition energies of 1.55 and 2.93~eV, lying nearby, corresponds to lasing at 428~nm. The scheme of level couplings is shown in Fig.~\ref{fig1}.  

The temporal evolution of this five-level system in a given laser electric field is described by a system of Bloch equations for the density matrix involving five diagonal elements representing the corresponding population of states A(2), A(3), B(0), X(0) and X(1), and six off-diagonal elements accounting for polarization couplings between A-X, B-X and A-B states as shown in Fig.~\ref{fig1}. The governing equations are presented in Appendix~\ref{app1}.

\begin{figure}[!ht]
\centerline{\includegraphics[width=7cm]{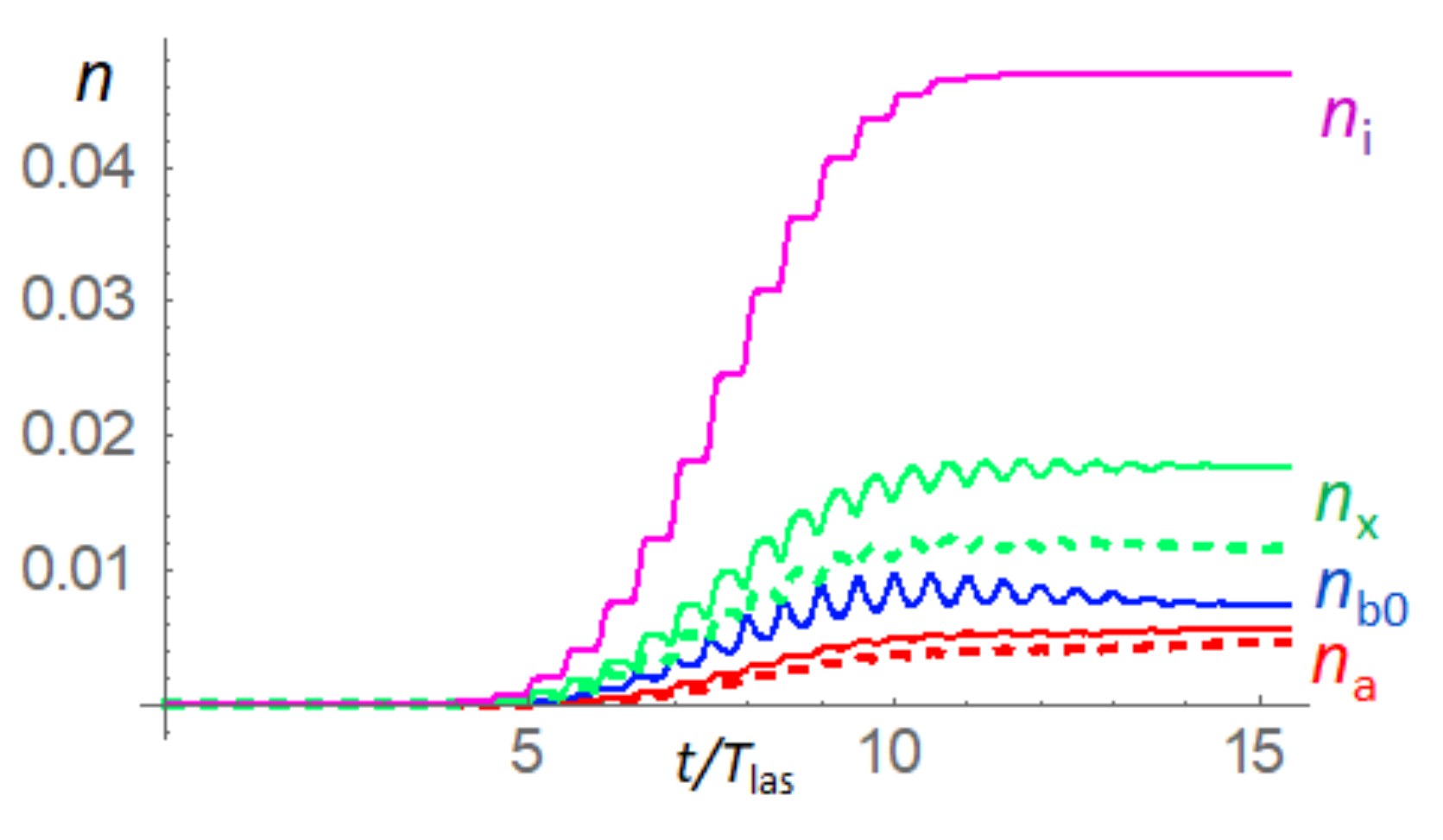} }
\caption{Time dependence of the ionization and excitation probabilities of nitrogen molecular ions created in a laser pulse with maximum intensity $I_{\rm las}=1.5\times10^{14}$\,W/cm$^2$, wavelength 800~nm and duration $t_{\rm las}=40$~fs for the molecule orientation at angle $\theta_n=18^\circ$ with respect to the laser electric field, $T_{\rm las}=2\pi/\omega_0=2.67$~fs is the laser period: the total ionization fraction -- pink line, population at X levels -- green lines, population at B level -- blue line, and population at A levels -- red lines. Solid lines correspond to the vibrational levels X(0) and A(2), dashed lines -- to the vibrational levels X(1) and A(3).} \label{fig2}
\end{figure}

Equations~\eqref{eqa11} -- \eqref{eqa62} were solved numerically for a given laser pulse maximum intensity $I_{\rm las}$ and pulse duration $t_{\rm las}$. The ion state probabilities were evaluated at the end of the laser pulse, $t=t_{\rm las}$, and averaged over the angle $\theta_n$ between the laser field and molecular axis by taking into account the angular dependence of the ionization probability $g(\theta_n)$ and the polarization couplings. We assume an isotropic distribution of molecules in the gas before the laser pulse arrival. Fig.~\ref{fig2} shows an example of temporal evolution of the state probabilities for the triplets X(0)-A(2)-B(0) (solid lines) and X(1)-A(3)-B(0) (dashed lines) for the molecule orientation of $18^\circ$ with respect to the laser polarization. Steps in the ionization curve correspond to the maxima of laser electric field and oscillations in the population levels are due to the polarization coupling. Ionization to the ground levels X is the dominant channel, the population at level X(0) is larger than at X(1). Direct ionization in excited states is a minor effect. The excited states A and B are populated essentially from levels X by the strong-field Rabi coupling. 

\begin{figure}[!ht]
\centerline{\includegraphics[width=15cm]{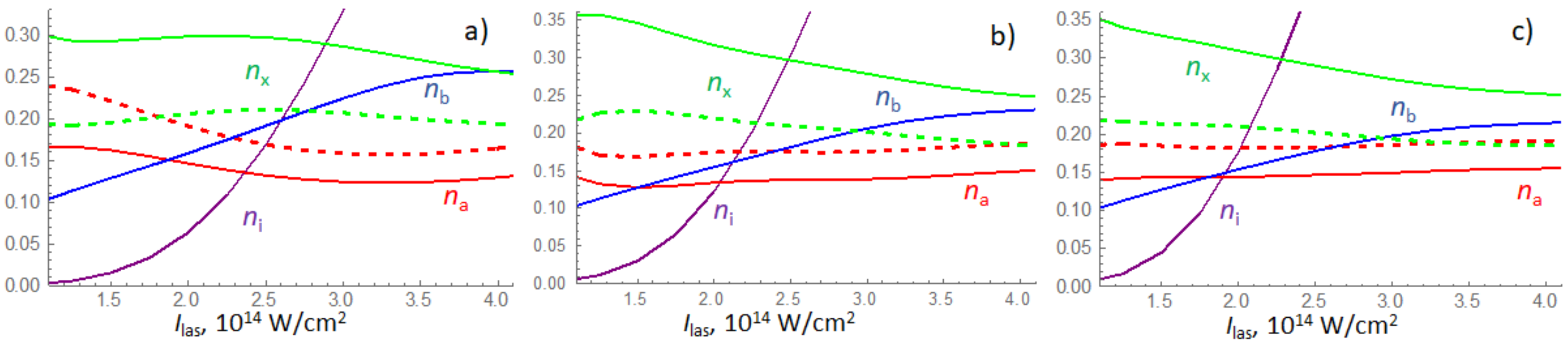} }
\caption{Dependence of the populations at levels A(2) (red), A(3) (red dashed), B(0) (blue), X(0) (green) and X(1) (green dashed) normalized to the total ion density, $n_i$ on the laser intensity for laser pulse duration 20~fs (a), 40~fs (b) and 60~fs (c). The total ionization probability $n_i/n_0$ is shown with purple line.} \label{fig3}
\end{figure}

Fig.~\ref{fig3} shows the dependence of the populations on laser intensity for laser pulse duration of 20, 40 and 60~fs. The calculated values are averaged over the molecular orientation assuming isotropic molecular distribution before the laser arrival. The fraction of ionized nitrogen molecules shown with purple lines increases with the laser pulse duration and intensity. Partition of ions between the excited levels depends essentially on the laser intensity and much less on the pulse duration. At the lowest intensities excitation to the level A(3) has the highest probability while excitation to the highest level B has the lowest probability. However, with the increase of laser intensity above $\sim2\times10^{14}$\,W/cm$^2$ the population ratio between the levels A and B inverses: Population at level B increases with laser intensity, while population at level A decreases.

This general trend presented in Fig.~\ref{fig3} is rather similar for the three pulse duration chosen in panels a, b, and c. Population at level X(0) is always the largest, while an inversion between states occurs at a laser intensity $\gtrsim 3\times10^{14}$~W/cm$^2$. The inversion between levels B and A(2) occurs at a laser intensity less than $2\times10^{14}$~W/cm$^2$, and the inversion point slightly varies with the pulse duration. By increasing the laser intensity beyond $4\times10^{14}$~W/cm$^2$ one may also achieve a population inversion between B and X(0) levels, but it is observed only with the shortest pulse duration and it corresponds to an ionization level of more than 40\%.  

The ionization-excitation model of nitrogen molecules with a short and intense laser pulse presented in this section is quite robust. The calculated populations depend rather weakly on exact values of the frequencies of B-X and A-X transitions and their detuning with respect to the laser frequency. Coupling is strong and it is controlled essentially by the large values of the Rabi frequencies $\mu_{a,bx}E_{\rm las}/\hbar$, comparable to the transition frequencies. Direct ionization to the excited states A(2) and A(3) accounted for in Eqs.~\eqref{eqa21} and \eqref{eqa22} makes a relatively small contribution of about 10\% to the population in the excited states. Contribution of the direct ionization to the level B(0) is even smaller, it is less than 1\% in the considered cases.

In the next section we investigate the evolution of the excited molecules in a gas after the end of the main laser pulse, assuming that the polarization corresponding to the A-X transition is maintained by a weak laser post-pulse. It is supposed that such a post-pulse cannot affect the population distribution between the levels but it is sufficiently strong and resonant for maintaining one of polarizations $d_{a2x0}$ or $d_{a3x1}$ for a time of a few ps, much longer than the main laser pulse duration. Due to the polarization coupling, the laser post-pulse may induces a delayed emission from the level B(0).

\section{Temporal evolution of the ion populations and seed amplification in the three-level interaction} \label{sec3}
\subsection{Maxwell-Bloch equations}\label{sec31}
Here, we consider the temporal and spatial evolution of the triplet X(0)-A(2)-B(0). One has to account for the possible evolution of the post-pulse in space and in time while propagating through the plasma filament. This implies the use of Maxwell-Bloch equations that account for the evolution of both ion populations and electromagnetic fields in the plasma. The post-pulse intensity is about four orders of magnitude smaller than the main pulse. It cannot ionize the gas and the coupling to molecular ionic transitions is weak. This allows us to treat each V-triplet independently, to use an envelope approximation for the electric fields and polarization fields and to select only the resonantly coupled levels. Consequently, the electromagnetic field is presented here as a sum of two components operating at the frequencies close to the transitions A(2)-X(0) and B(0)-X(0):
$$ E(z,t)= {\rm Re}\big[E_{ax}\exp(-i\omega_{ax} \tau) +E_{bx}\exp(-i\omega_{bx} \tau) \big],$$
where $\tau=t-z/c$  is the co-propagation time. Possible detuning from the resonant frequency is accounted for in the time dependence of the electric field amplitude $E_{a,bx}(\tau,z)$. Similarly, all polarizations in the Bloch equations~\eqref{eqa41} -- \eqref{eqa62} are separated in the slow varying amplitudes and fast oscillating phases:
$$ d_{ax}=-\frac i2 p_{ax}\exp(-i\omega_{ax}\tau), \qquad  d_{bx} =-\frac i2 p_{bx}\exp(-i\omega_{bx}\tau), \qquad d_{ba}=\frac12 p_{ba}\exp(-i\omega_{ba} \tau).$$
Then the Maxwell-Bloch equations for the three level system system read~\cite{Kozlov1999}:
\begin{eqnarray}
&&c\partial_z E_{ax}=\frac{\omega_{ax} n_i \mu_{ax}}{2\epsilon_0} p_{ax} \sin\theta_i, \qquad c\partial_z E_{bx}=\frac{\omega_{bx} n_i \mu_{bx}}{2\epsilon_0} p_{bx}\cos\theta_i, \label{eq7e}\\
&& \partial_\tau  p_{ax}= -\gamma_{ax} p_{ax}+ \frac{\mu_{ax}}{\hbar} E_{ax} (n_a-n_x) \sin\theta_i +\frac{ \mu_{bx}}{2 \hbar} E_{bx} p_{ba}^*  \cos\theta_i, \label{eq7}\\
&& \partial_\tau p_{bx}= -\gamma_{bx} p_{bx}+\frac{\mu_{bx}}{\hbar} E_{bx} (n_b-n_x) \cos\theta_i+\frac{  \mu_{ax}}{2 \hbar} E_{ax} p_{ba}  \sin\theta_i, \label{eq8} \\
&&  \partial_\tau p_{ba}=  -\gamma_{ba} p_{ba}-\frac{\mu_{bx}}{2 \hbar}E_{bx} p_{ax}^* \cos\theta_i -\frac{  \mu_{ax}}{2 \hbar} E_{ax}^* p_{bx}\sin\theta_i,   \label{eq9} \\
&& \partial_\tau n_a= -  \frac{\mu_{ax} }{2\hbar} {\rm Re}(p_{ax}^*E_{ax})\, \sin\theta_i, \qquad \partial_\tau n_b= -  \frac{\mu_{bx} }{2\hbar} {\rm Re}(p_{bx}^*E_{bx})\, \cos\theta_i,   \label{eq10}
\end{eqnarray}
where $\theta_i$ is the angle of the ion molecule orientation with respect to the laser polarization, coefficients $\gamma_{ij}$ account for the collision-induced spontaneous damping of the corresponding polarizations, $n_i$ is the density of nitrogen ions in the plasma and the populations of all three levels are normalized to the ion density and are related by the condition $n_{a}+n_{b}+n_{x}=1$. The characteristic time of spontaneous damping is on the order of a few ps, which is comparable to the amplification time, and needs to be retained in these equations. Initial conditions for this system are provided by solution of the Bloch equations~\eqref{eqa11} -- \eqref{eqa62} driven by the main pulse.

It is convenient for the qualitative analysis and numerical solutions to introduce the characteristic time $t_N=(\epsilon_0\hbar/\omega_{ax} \mu_{ax}^2 n_i)^{1/2}$ and the characteristic length $z_N=ct_N$. In what follows we shall use the dimensionless time $\tau\to\tau t_N$ and coordinate $z\to z/ct_N$ and normalized electromagnetic fields $e_{a,b}=\mu_{a,bx} E_{a,bx}t_N/\hbar$. (Numerical values for these
parameters are given in Sec.~\ref{sec4}.) Then the dimensionless set of equations can be cast in the following form
\begin{eqnarray}
&&\partial_z e_a=\frac12 p_{ax} \sin\theta_i, \qquad    \partial_\tau  p_{ax}=  -\hat\gamma_{ax} p_{ax}+(n_a-n_x)\, e_a\sin\theta_i +\frac{1}{2}  p_{ba}^* e_b \cos\theta_i, \label{eq11}\\
&&\partial_z e_b=\frac{r}2  p_{bx}\cos\theta_i, \qquad   \partial_\tau p_{bx}= -\hat\gamma_{bx} p_{bx}+ (n_b-n_x)\, e_b \cos\theta_i +\frac{1}{2}  p_{ba} e_a \sin\theta_i,\label{eq12}\\
&& \partial_\tau n_a= -\frac12 {\rm Re}(p_{ax}^*e_a)\,\sin\theta_i, \qquad  \partial_\tau n_b= -\frac12{\rm Re}(p_{bx}^*e_b)\,\cos\theta_i , \label{eq131}\\
&& \partial_\tau p_{ba}= -\hat\gamma_{ba} p_{ba} -\frac{1}{2}  p_{ax}^*e_b\cos\theta_i -\frac{1}{2}  p_{bx} e_a^*\sin\theta_i.\label{eq132}
\end{eqnarray}
where $r=\omega_{bx}\mu_{bx}^2/\omega_{ax}\mu_{ax}^2$ is the ratio of characteristic times of evolution of the B-X and A-X transitions and $\hat\gamma_{ij}= \gamma_{ij}t_N$ are the dimensionless damping rates. For the nitrogen ion this ratio  $r\simeq18$ is quite large because of a large ratio of dipole moments. This implies a much faster temporal evolution of the B-X transition compared to A-X.

It is important to mention that in the absence of damping, $\hat\gamma_{ij}=0$, this system has the propriety of conserving locally in space the following combination of populations:
\begin{equation} \label{eq14}
\frac12|p_{ax}|^2+ \frac12|p_{bx}|^2+ \frac12|p_{ba}|^2 +n_a^2 + n_b^2 + n_x^2=C,
\end{equation} 
where $C$ is a positive constant which is equal to $1/3$ for fully decorrelated and equally partitioned populations. The case of fully correlated system $C=1$ is considered in Sec.~\ref{sec333}. Another important property is the energy conservation in the system. By time integrating the equations for $|e_{a,b}|^2$, one obtains the energy conservation laws for transitions A-X and B-X:
\begin{eqnarray} 
&& F_{a}(L) -  F_{a}(0) =\int_0^L dz\, \left(n_{a}(z,0) - n_{a}(z,\tau_{\max})\right), \label{eq10a}\\
&&  F_{b}(L) -  F_{b}(0) =\int_0^L dz\, \left(n_{b}(z,0) - n_{b}(z,\tau_{\max})\right). \label{eq10b}
\end{eqnarray}
Here $F_a$ and $F_b$ are the energy fluxes of the pump and seed pulses at the end of simulation $\tau=\tau_{\max}$ at the entrance, $z=0$, and the exit, $z=L$, of the system
$$ F_{a}(z)= \frac12\int_0^{\tau_{\max}} d\tau\,|e_{a}(z,\tau)|^2, \qquad
F_{b}(z)=\frac1{2r}\int_0^{\tau_{\max}} d\tau\,|e_{b}(z,\tau)|^2.$$
These relations~\eqref{eq10a} and \eqref{eq10b} confirm that the number of emitted photons in A-X and B-X transition is conserved separately and it is equal to the number of ions transferred from the correspondent excited state to the ground state. By taking the difference between the equations for the field intensities one can obtain also the following relation
$$  \frac12 \partial_z|e_{a}|^2 -\frac{1}{2r}\partial_z |e_{b}|^2 = \partial_\tau n_{b} -  \partial_\tau n_{a}, $$
which relates the number of ions transferred from state B to state A through the ground state X to the number of emitted photons in the B-X and A-X transitions. This corresponds to a lasing process without population inversion but has also been viewed as a two-photon stimulated Raman scattering~\cite{Breddermann2018}. However, in difference from the conventional, single photon Raman scattering, here the scattered wave is up-shifted in frequency and the energy is provided from the medium. Therefore, in our view, it is better described as a lasing process without population inversion.

Fulfillment of the conservation laws \eqref{eq14}, \eqref{eq10a} and \eqref{eq10b} provides a test for the accuracy of numerical calculations shown below. The positive values of $F_{a,b}$ correspond to the amplification of the corresponding input signal in the plasma. In what follows, we define the energy gain, $G_{a,b}$, as a ratio between the emitted flux at the end of simulation and the injected flux,
and $\eta_b$ as the fraction of energy extracted from the excited level:
\begin{equation} \label{eq10c}
G_{a,b}=F_{a,b}(L)/F_{a,b}(0), \qquad  \eta_b=F_b(L)/n_{b0}L.
\end{equation}
Before discussing numerical solutions we first present several analytic results that are useful for their interpretation. 

\subsection{Analytical solutions to the three level system} \label{sec33}
The system of equations describing the V-scheme has several analytical solutions that are presented in Appendix~\ref{app2} for reference. Here, we present two particular solutions that will help us interpret the numerical simulations presented in the next section: the lasing without inversion and the solitons. In both cases we neglect the spontaneous damping for sake of simplicity.

\subsubsection{Lasing without population inversion} \label{sec334}
The system of equations~\eqref{eq11} -- \eqref{eq132} has a particular solution, discussed by Svidzinsky et al.~\cite{Scully2013}. It corresponds to an amplification of a signal at a higher frequency $\omega_{bx}$ due to a parametric coupling to the pump wave of lower frequency $\omega_{ax}$ through the correlated polarization between levels A and B. Here, we assume that there are populations at both excited states, $n_{a0}$ and $n_{b0}$, and that the A-X transition is driven by a pump field $e_a (\tau,z)=e_{a0}\exp(-i\Delta\omega_{ax}\tau+iqz)$ coupled to the polarization $p_{ax} (\tau,z)=p_{ax0}\exp(-i\Delta\omega_{ax}\tau+iqz)$. Relations between the amplitudes, $p_{ax0}=2iqe_{a0}/\sin\theta_i$, the space dephasing parameter $q=(n_{a0}-n_{x0})\sin^2\theta_i/2\Delta\omega_{ax}$ and the frequency detuning $\Delta\omega_{ax}$ follow from the first two equations of the system.
 
Let us consider seed wave of a small amplitude $e_{b0}$ at a frequency $\omega_{bx}$ corresponding to the B-X transition and investigate the linear response of the system~\eqref{eq11} -- \eqref{eq132} to this initial perturbation. The perturbed system contains three equations for the field $e_b$ and polarizations $p_{bx}$ and $p_{ba}$:
\begin{eqnarray}
&& \partial_z e_{b}=\frac{r}2 p_{bx}\cos\theta_i, \qquad  \partial_\tau p_{bx}= (n_{b0}-n_{x0} ) e_{b}\cos\theta_i +\frac{1}2 p_{ba} e_{a0}\sin\theta_i\exp(-i\Delta\omega_{ax}\tau +iqz), \nonumber \\
&& \partial_\tau p_{ba}= iq e_{a0}e_b\cos\theta_i \exp(i\Delta\omega_{ax}\tau-iqz) -\frac{1}2 p_{bx} e_{a0}\sin\theta_i \exp(i\Delta\omega_{ax}\tau-iqz). \nonumber
\end{eqnarray}
Making a Fourier transform in time with the initial condition $e_b (0,z)=e_{b0}$, and expressing the Fourier components of polarizations through $e_b$, one finds an equation for the seed field $\partial_z e_b=H\,e_b$ with the gain factor $H$ depending on detuning:
$$ H=\frac{ir\cos^2\theta_i}{2\Delta\omega_{ax}} \,  \frac{\Delta\omega_{ax} (\omega -\Delta\omega_{ax} )\,(n_{b0}-n_{x0})-(n_{a0}-n_{x0})\,e_{a0}^2\sin^2\theta_i/4}{\omega\,(\omega- \Delta\omega_{ax})-e_{a0}^2\sin^2\theta_i/4}.$$
Solving this equation, the signal amplitude can be expressed as an inverse Fourier transform:
\begin{equation}\label{eq17}
e_b(\tau,z)=\frac{ie_{b0}}{2\pi} \int \frac{d\omega}\omega \exp(-i\omega\tau+Hz).
\end{equation}
The integral here is performed in the complex plane $\omega$ along the contour going above the singular points, according to the principle of causality. The singular points are solutions of the dispersion equation: $$\omega\,(\omega-\Delta\omega_{ax})-e_{a0}^2\sin^2\theta_i/4 =0,$$ 
which has two solutions $\omega_{1,2}=\frac12 \Delta\omega_{ax}\mp \frac12\sqrt{\Delta\omega_{ax}^2+ e_{a0}^2\sin^2\theta_i}$. The first one corresponds to a spontaneous amplification at the B-X transition modified by the presence of the A level. It requires the population inversion $n_{b0}>n_{x0}$ as it is shown in Appendix~\ref{app2}b. The second one corresponds to a parametric coupling between A and B levels and requires a softer condition  $n_{b0}>n_{a0}$.  A convenient way to compute the inverse Fourier transform~\eqref{eq17} asymptotically in the limit $rz\tau\cos^2\theta_i \gg1$ is to close the integration contour in the lower half plane and to transform it into two circles with the centers in the singular points $\omega_{1,2}$. The radius of each circle is found by equating the amplitudes of the two terms in the exponential. In particular, assuming a sufficiently large detuning $ |\Delta\omega_{ax}|\gg e_{a0}|\sin\theta_i|$, for the singular point $\omega_1\approx -e_{a0}^2\sin^2\theta_i/4 \Delta\omega_{ax}$, the circle radius $\Gamma_1$ is defined by equation: $2\Gamma_1^2 \tau \approx rz(n_{b0}-n_{x0})\cos^2\theta_i$. This singular point corresponds to oscillations decaying in time with the characteristic frequency $\Gamma_1$, if $n_{b0}<n_{x0}$, or to an exponential growth in time, if $n_{b0}>n_{x0}$. 

The radius of the contour around the second singular point  $\omega_2\approx \Delta\omega_{ax}+e_{a0}^2\sin^2\theta_i/4 \Delta\omega_{ax}$ is defined by equation: $8\Gamma_2^2 \tau\approx rz(n_{b0}-n_{a0}) e_{a0}^2 \cos^2\theta_i\, \sin^2\theta_i/ \Delta\omega_{ax}^2$. The integral around this singular point corresponds to the modified Bessel function $I_1$ in the case $n_{b0}>n_{a0}$: 
\begin{equation}\label{eq19}
e_b (\tau,z)=e_{b0}  \frac{\Gamma_2}{|\Delta\omega_{ax}|}  I_1 (2\Gamma_2 \tau).
\end{equation}
In the limit $2\Gamma_2 \tau\gg 1$ this expression corresponds to an exponentially growing solution. In physical units expression for the gain factor reads
\begin{equation}\label{eq20}
  G_{bx}\simeq\frac{\mu_{bx} E_{a0}}{2\hbar}|\sin2\theta_i|  \frac{\sqrt{n_{b0}-n_{a0}}}{|\Delta\omega_{ax} | t_N} \sqrt{\frac{\omega_{bx} \tau z}{2\omega_{ax}c}}.
\end{equation}
If spontaneous damping is included, it imposes a threshold value of the pump field amplitude for excitation of this instability. 

\subsubsection{Solitary excitation} \label{sec333}
The soliton solution is a dynamic structure that propagates along the plasma with a constant velocity. It was described in a two-level system by McCall and Hahn~\cite{McCall1967} and in more details in Ref.~\cite{Newell1992}. Considering for example A-X transition alone and neglecting the polarization damping term, system~\eqref{eq11} and \eqref{eq131} reduces to the following three equations:
\begin{equation}\label{eq15}
\partial_z e_a=\frac12p_{ax}\sin\theta_i, \qquad
\partial_\tau  p_{ax}= (n_a-n_x) e_a \sin\theta_i, \qquad  \partial_\tau n_a= - \frac12{\rm Re}(p_{ax}^*e_a)\,\sin\theta_i.
\end{equation}
Assuming $e_a$ to be real and depending on the coordinate and time as $\xi=(\tau-z/u)\,\sin\theta_i$, this system has two integral relations, $|p_{ax}|^2=2 n_a (1-n_a)$ and $e_a^2=2n_a u$, which are the particular case of the more general expressions~\eqref{eq14} and \eqref{eq10a}. The constants in these two relations are chosen assuming that there is no population at the level A before the laser pulse arrival. Then the remaining equation for $n_a$ has a soliton solution:
\begin{equation}\label{eq16}
n_a=\frac{1}{\cosh^2(w\xi)}, \qquad e_a=\frac{e_0}{\cosh(w\xi)}, \qquad p _{ax}=-\frac{\sinh(w\xi)}{\cosh^2(w\xi)},
\end{equation}
where the amplitude $e_0$, velocity $u$ and inverse width $w$ are related as $e_0=\sqrt{2u}$ and $w=e_0/2\sqrt{2}$. Population at the level A increases from 0 to 1 when $\xi$ is negative and increasing, while it decreases back to zero when $\xi$ is positive. The soliton carries an area $A=  \int_{-\infty}^\infty e_a d\xi=2\sqrt{2}\pi$. That relation imposes a condition on the minimum pump amplitude needed for the soliton excitation.

The soliton propagates with a velocity $uc/(1+u)$ smaller than the light velocity. It transports a dimensionless energy flux $F_a=\frac12 \int_{-\infty}^\infty e_a^2dz=e_0^3\sqrt{2}/\sin\theta_i$, which depends on its amplitude. Soliton solutions for a three level system has been constructed in Refs.~\cite{Hioe1994, Chakravarty2016} for a particular case of $r=1$. However, these are not relevant to our conditions of the large value of $r\approx 18$ that makes correlation between B-X and A-X solitons more complicated. Particular solutions of the V-system described above are found in the numerical analysis presented in the next section. 

\section{Numerical solutions for the V-system} \label{sec4}
\subsection{Initial conditions} \label{sec40}
Before discussing the numerical solutions for the set of equations~\eqref{eq11} -- \eqref{eq132}, we need to consider the role of molecular rotations. The strong laser electric field induces a dipolar moment in a neutral nitrogen molecule and exerts a torque. This leads, after an inertial delay of $\sim100$~fs, to the formation of a coherent rotational wavepacket with a partial alignment of the neutral molecules along the laser field axis. As molecules have a broad discrete distribution in the rotational moments, a coherent rotational wave packet quickly dephases but then experiences spontaneous revivals every half rotation period $T_{\rm rot} = 1/(2Bc)$~\cite{Ripoche1997, Miyazaki2005}. (Here, $B$ is the rotation constant equal to 2.0~cm$^{-1}$ for the neutral molecule, 2.07~cm$^{-1}$ for level B and 1.93~cm$^{-1}$ for level X.) The duration of revivals is rather short, it is $J_0$ times shorter than the revival period, where $J_0 \simeq\mu E_{\rm las}t_{\rm las}$ is the characteristic rotation momentum.

For the parameters of interest in our study $J_0 \sim 10-20$ and the corresponding revival duration is less than 1~ps. Thus, revivals should not affect significantly the amplification process that proceeds on a longer time scale. Therefore, in our model we assume that the probability of angular distribution of ions, $\mathcal{P}(\theta_i)$, does not depend on time, and we consider a quasi-classical angular distribution of ions with the average value $\langle\cos^2\theta_i\rangle\simeq 0.33$ corresponding to a non-adiabatic strong short pulse excitation~\cite{ Miyazaki2005, Zeng2009}. Equations for angle averaged populations, $n_{a,b,x}$, and corresponding polarizations are given in Appendix~\ref{app3}.

In the numerical analysis of our system we consider the B(0)-A(2)-X(0) transition at wavelengths 391.4 and 787.5~nm~\cite{Liu19}. The fractions of excited ions and initial values of polarizations are calculated from the system of Bloch equations discussed in Sec.~\ref{sec2}. As an example, we consider the main laser pulse intensity $2.6\times10^{14}$~W/cm$^2$ and duration of 20~fs. According to Fig.~\ref{fig3}, that choice of parameters corresponds to an ionization of 20\% and to a situation without population inversion with respect to the ground level X: $n_{a0}=0.12$, $n_{b0}=0.20$ and $n_{a0}=0.30$. However, it satisfies the necessary amplification condition $n_{a0}<n_{b0}$.  

We first consider solutions of system~\eqref{eq11} - \eqref{eq132} without external fields, $e_{ax}=e_{bx}=0$, and with maximum initial values of polarizations: $p_{ij}\sim 1$. (The choice of phases has no importance.) The system with these initial conditions is quickly discharged by spontaneously amplifying photons at both transitions.  The characteristic de-excitation times of the system, $\Delta t_a\sim ct_N^2/(2L|n_a-n_x |)$  and $\Delta t_b \sim ct_N^2/(2rL|n_b-n_x |)$  are very short, less than $0.1~t_N$, especially for the B-X transition. This is consistent with the conservation equation~\eqref{eq14}: any state will terminate with zero polarizations at excited levels and with $n_x=1$. This situation, however, is not consistent with the delayed emission observed in the experiments. By reducing polarization amplitudes by 100 times or more, one can slow down the spontaneous emission and maintain a large fraction of ions in the excited states. However, we verified that without feeding the polarization $p_{ax}$ with an external pump it is not possible to obtain an efficient coupling between levels A and B and amplification. Therefore, all simulations presented below were conducted with a post-pulse pump applied at $t=0$ and decaying exponentially with time.

\subsection{Reference case} \label{sec41}
As a reference point we take a nitrogen gas pressure $p_N=30$~mbar, which corresponds to a total ion density $n_i=1.2\times10^{17}$~cm$^{-3}$ for a 20\% ionization for the three considered states. Then, according to Sec.~\ref{sec31}, the characteristic time becomes $t_N \simeq 0.85$~ps, the characteristic length $ct_N \simeq 0.26$~mm, the effective electric field $E_N=\hbar/\mu_{ax}t_N \sim 58.5$~MV/m and the normalization intensity $I_N= c \epsilon_0 E_N^2=0.91$~GW/cm$^2$. Spontaneous decorrelation time $1/\gamma$ at this pressure is set to 8.5~ps, same for all three polarizations.

We consider a filament length of $L=7.8$~mm. The post-pulse  at wavelength 788~nm corresponding to detuning $\Delta\omega_{ax}=-3.5$~ps$^{-1}$ is decaying exponentially with a time constant of $t_a=5\,t_N=4.3$~ps. Its initial electric field $E_{a0}=293$~MV/m corresponds to the dimensionless amplitude $e_{a0}=5$ and intensity 11.4~GW/cm$^2$. The fluence $F_a (0)=49$~mJ/cm$^2$ injected with the post-pulse is significantly larger than the energy stored at the A level $\hbar\omega_{ax} n_{a0} n_i L=4.7$~mJ/cm$^2$. 

A seed pulse at a frequency corresponding to the B-X transition and with a duration of 0.17~ps is injected 0.43~ps after the end of the main pump pulse. The seed field $E_{b0}=19.5$~MV/m corresponds to the dimensionless amplitude $e_{b0}=1$ and the intensity 0.05~GW/cm$^2$. The injected seed fluence $F_b (0)=0.009$~mJ/cm$^2$ is more than 3 orders of magnitude smaller than the energy stored in the B level, $\hbar\omega_{bx} n_{b0} N_i L=15.3$~mJ/cm$^2$. The choice of seed amplitude has not much importance. The initial polarizations of the order of $10^{-2}-10^{-3}$ are taken from the solution of the Bloch system discussed in Sec.~\ref{sec2}. Their phases are not important.It is also possible to obtain emission from the B-X transition with other seed amplitudes or without seed provided there is small initial polarization $p_{bx0}$ in the filament. Such an initial polarization $p_{bx0}$ by field induced recollisions has been attributed as the source for the amplification in the absence of a seed pulse~\cite{Liu15, Liu2017}.  

\begin{figure}[!ht]
\centerline{\includegraphics[width=12cm]{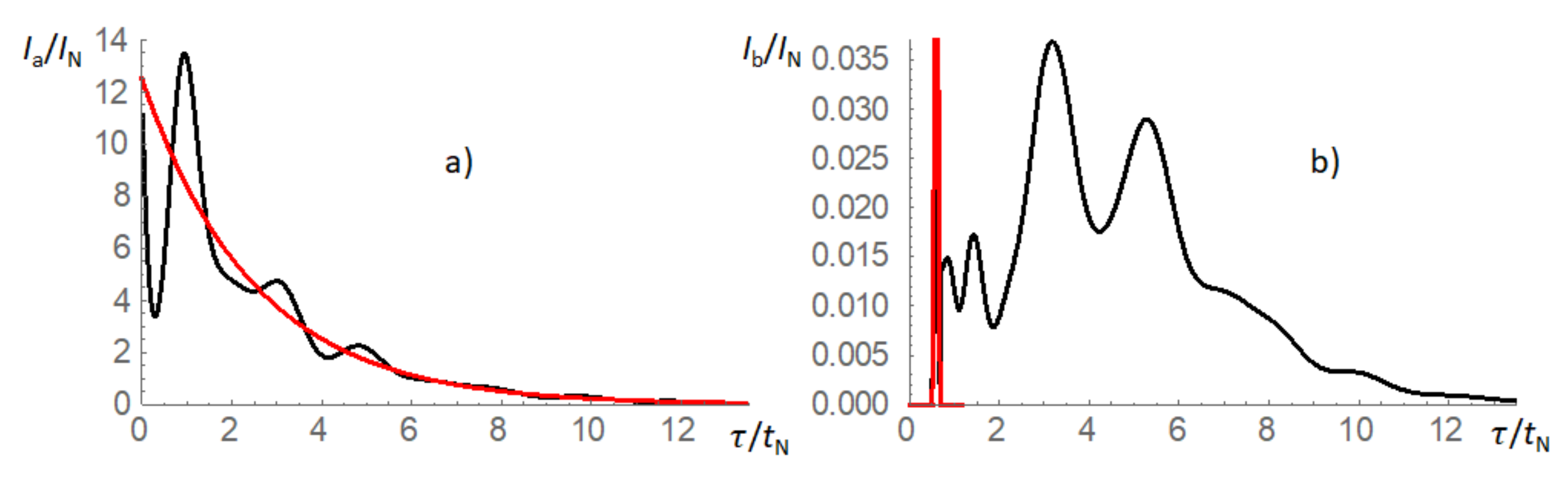} }
\caption{Temporal evolution of the intensity of electromagnetic waves $I_a$ (a) and $I_b$ (b). Red curves -- incident field, $I_{a,b} (0,t)$, black lines -- exiting field, $I_{a,b} (L,\tau)$. The intensity and time are in the normalized units: $t_N=0.85$~ps and $I_N=0.91$~GW/cm$^2$. Dimensionless pump amplitude $e_{a0} = 5$, other conditions and parameters are given in the text.}  \label{fig4}
\end{figure} 

Example of the seed amplification in these conditions is presented in Fig.~\ref{fig4}. It shows the intensities of the pump and seed pulses, $I_a/I_N=\frac12 |e_a |^2$ and $I_b/I_N=\frac12 |e_b|^2 (\mu_{ax}/\mu_{bx})^2$, at the entrance of the filament, $z=0$, and at the exit, $z=L$, in function of co-propagation time $\tau$. Here $I_N=\epsilon_0 cE_N^2=0.91$~GW/cm$^2$ is the normalization intensity. The pump wave is modulated at the exit with a period of $(2-3)\,t_N$ due to the frequency detuning from the A-X transition and partial absorption. The seed pulse of duration $0.2\,t_N$ is injected at time $0.5\,t_N$ with a duration $0.25\,t_N$. It is amplified more than 25 times in energy and it extracts about 0.8\% of the energy initially stored in level B. 

\begin{figure}[!ht]
\centerline{\includegraphics[width=12cm]{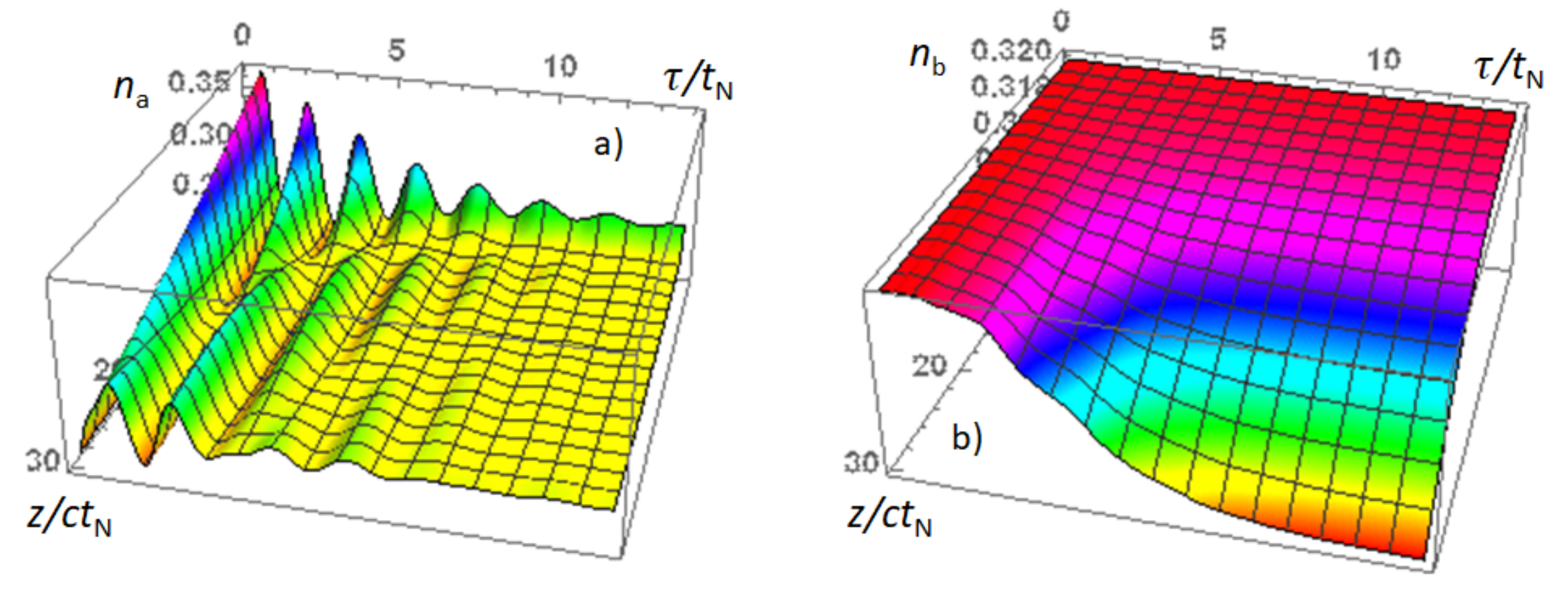} }
\caption{Spatio-temporal evolution of the populations at the excited level A (a) and B (b). Parameters are the same as in Fig.~\ref{fig4}.}  \label{fig5}
\end{figure} 

The amplified signal is extended by more than 10~ps and produces at the exit a sequence of pulses of duration of about 2~ps. The amplification is explained by the parametric coupling of levels A and B by the polarization $p_{ba}$ as it is described in Sec.~\ref{sec334}. By suppressing the corresponding cross-polarization terms in Eqs.~\eqref{eq31} -- \eqref{eq332} one may eliminate completely the amplification. It is also verified that  the gain is proportional to the difference of populations between levels B and A,  $n_{b0}- n_{a0}$, the filament length, $L$, and the pump amplitude, $e_{a0}$, according to  Eq.~\eqref{eq20}. In particular, no gain is found for equal populations, or when  $n_{a0}\geq n_{b0}$. 

The cross-coupling between levels A and B manifests itself also in the spatial and temporal evolution of the populations of excited ions shown in Fig.~\ref{fig5}. The propagation of the pump post-pulse induces Rabi oscillations of populations at the levels A and X with an amplitude decreasing with time. Part of the pump pulse is absorbed and the population at level A temporally increases up to $n_a\simeq 0.3$. It is accompanied by a corresponding decrease of the population at level X. By contrast, the population at level B shows a delayed decrease corresponding to an exponential increase of the signal amplitude. Nevertheless, the condition of amplification, $n_{a0}<n_{b0}$ is fulfilled all the time.

\subsection{Pressure dependence of signal gain} \label{sec42}
\begin{figure}[!bt]
\centerline{\includegraphics[width=15cm]{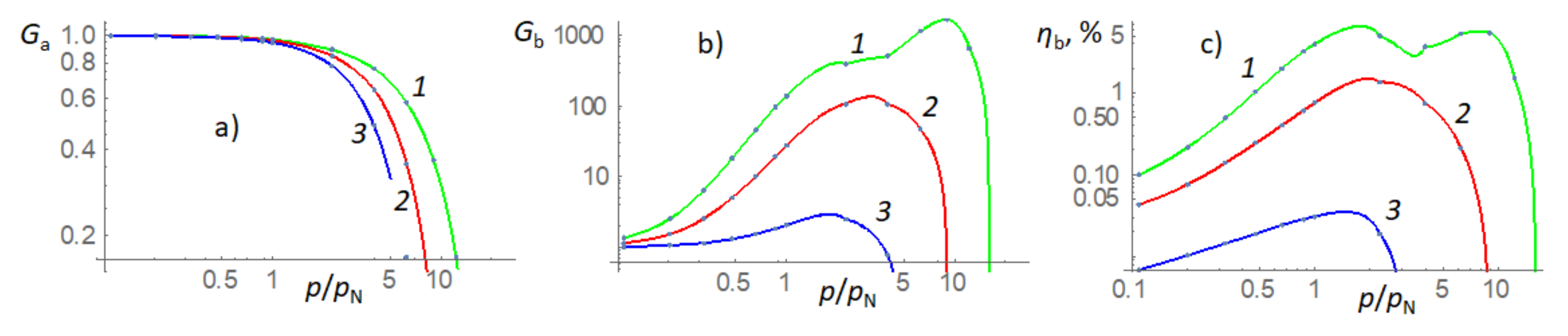} }
\caption{Pressure dependence of the energy gain of the pump post-pulse $G_a$ (a) and the seed pulse $G_b$ (b) and fraction of energy extracted from B-level $\eta_b$ in per cent (c). The nominal pressure $p_N=30$~mbar corresponds to the particle density $n_i=10^{17}$~cm$^{-3}$ on the three considered levels. The input pump intensity $I_a (0)=22.5$~GW/cm$^2$ ($e_{a0}=7$, green, 1), 11.5~GW/cm$^2$ ($e_{a0}=5$, red, 2), 4.1~GW/cm$^2$ ($e_{a0}=3$, blue, 3), the frequency detuning $\Delta\omega_{ax}=-3.5$~ps$^{-1}$, the post-pulse decay time is 4.3~ps. The input seed intensity $I_b (0)=0.05$~GW/cm$^2$, zero detuning, switched on at $t_b=0.5\,t_N$ for duration of $0.2\,t_N$, $n_{a0}=0.2$ and $n_{b0}=0.32$. The filament length $L=7.8$~mm.}\label{fig6}
\end{figure}

The parametric dependence of the amplification process is analyzed by scanning the gains of the pump and the seed pulses as a function of gas pressure and pump amplitude, while keeping all other parameters unchanged, except the spontaneous damping which varied linearly with pressure, $\gamma =\gamma_0 p/p_N$, where $\gamma_0=0.1/t_N$ has the same value as in the previous section. We assume that the ionization level remains the same and the density of ionized molecules is proportional to the pressure. Figure~\ref{fig6} shows variation of energy gain of the pump post-pulse, $G_a$, and seed pulse, $G_b$, as well as the fraction of energy extracted from level B, $\eta_b$, as a function of gas pressure.

\begin{figure}[!bt]
\centerline{\includegraphics[width=15cm]{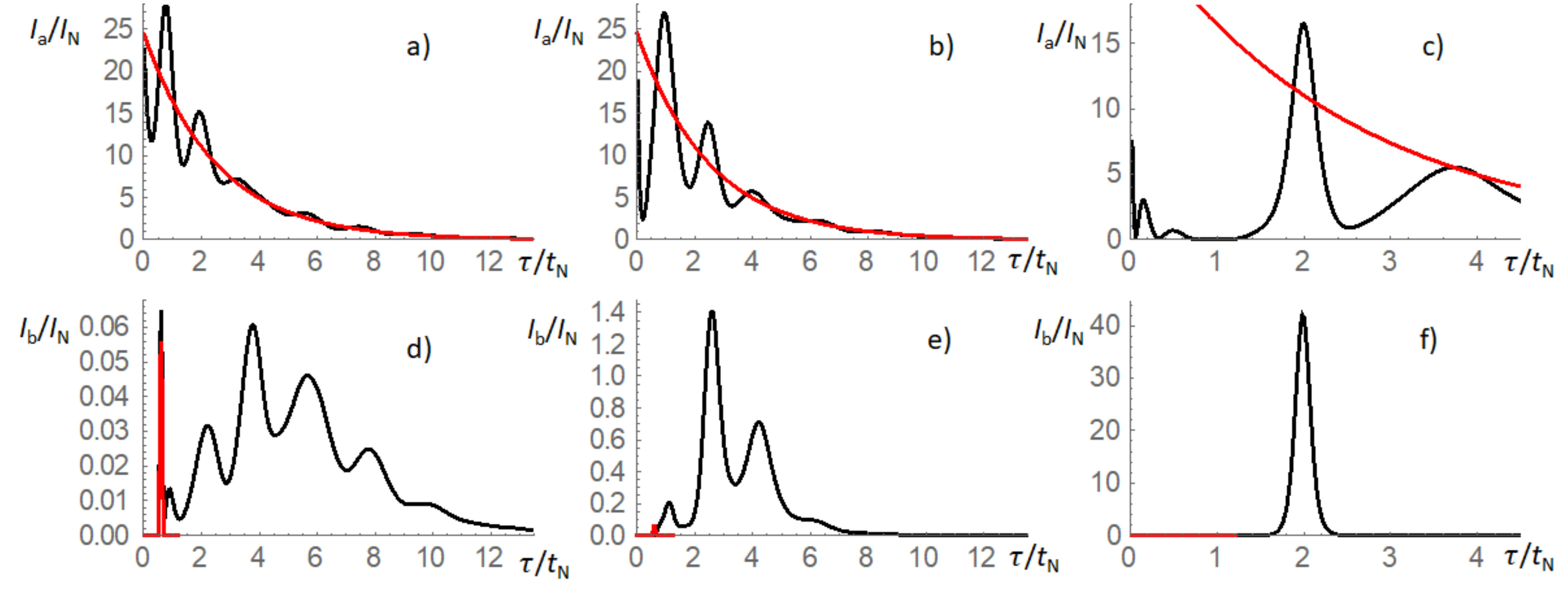} }
\caption{Temporal shape of the post-pump (a, b, c) and seed pulse (d, e, f) for gas pressures $p=0.81\,p_N$ (a), $p=2.25\,p_N$ (b) and $p=9\,p_N$ (c). The dimensionless pump amplitude $e_{ea0} = 7$, other parameters are the same as in Fig.~\ref{fig4}. The intensity and time and normalized to the values at the nominal pressure: $I_N=0.91$~GW/cm$^2$ and $t_N=0.85$~ps. Red lines show the injected pump and seed pulse.}  \label{fig7}
\end{figure}

By changing pressure by a factor of 300, from $0.1\,p_N$ to $30~p_N$, we observe two different behaviors at low and high pressures. At low pressures, the pump post-pulse goes through the plasma without any significant depletion, and its temporal shape is modulated at the Rabi frequency, see Figs.~\ref{fig4}a and \ref{fig7}a. By contrast, at high pressures, $p/p_N\gtrsim 3-10$, a significant part of the pump post-pulse is absorbed and the remaining part comes out with a $2-3$~ps delay in a form of the soliton, see Fig.~\ref{fig7}c.

The seed gain increases monotonously in power approximately $2.5-3$ at low pressures, where there is no pump pulse depletion, see curves 1 and 2 in Fig.~\ref{fig6}b. Such an increase of amplification with pressure follows directly from Eq.~\eqref{eq20}: the gain is proportional to the pump amplitude. This is the regime of parametric amplification. The amplified pulse is characterized by a long duration up (to 10~ps) accompanied by Rabi oscillations with a period $2-3$~ps. The amplification is, however, suppressed at very low pump amplitudes, see curve 3 in Fig.~\ref{fig6}b. In this case the signal growth rate is too low and it cannot be amplified during the time when the pump is present in plasma.
A cutoff at high pressures is explained by two effects: (i) the increase of spontaneous damping and (ii) post-pump absorption. For the chosen damping rate, the damping time is shorter than the amplification time  at pressures $p\gtrsim 10\,p_N$. As the parametric growth rate is proportional to the pump amplitude, the cutoff shifts to higher pressures if the pump is stronger. A second effect is demonstrated in Fig.~\ref{fig6}a: at pressures $p\gtrsim 5\,p_N$ the energy carried with the post-pulse is comparable with the energy stored in level A. Consequently, the pump is resonantly absorbed and cannot support the seed amplification. 

It is important to note that there is no correlation in the temporal shape of the transmitted pump post-pulse and seed pulse in the regime of parametric amplification. Conversely, at high pressures both the seed and the pump post-pulse exit the plasma synchronously with a time delay of $2-3$~ps and a sub-ps duration. This regime of soliton amplification is realized at pressures exceeding $(5 -10)\,p_N$, if the pump amplitude is sufficiently strong.

The increase of signal gain is correlated with an increase of the energy extracted from level B, $\eta_b$, in the pressure range corresponding to the parametric amplification regime. This is shown in Fig.~\ref{fig6}c in per cent. However, the extracted energy is saturated in the pressure range of $(3-10)\,p_N$, and in the soliton regime the extracted energy decreases. 

\begin{figure}[!ht]
\centerline{\includegraphics[width=12cm]{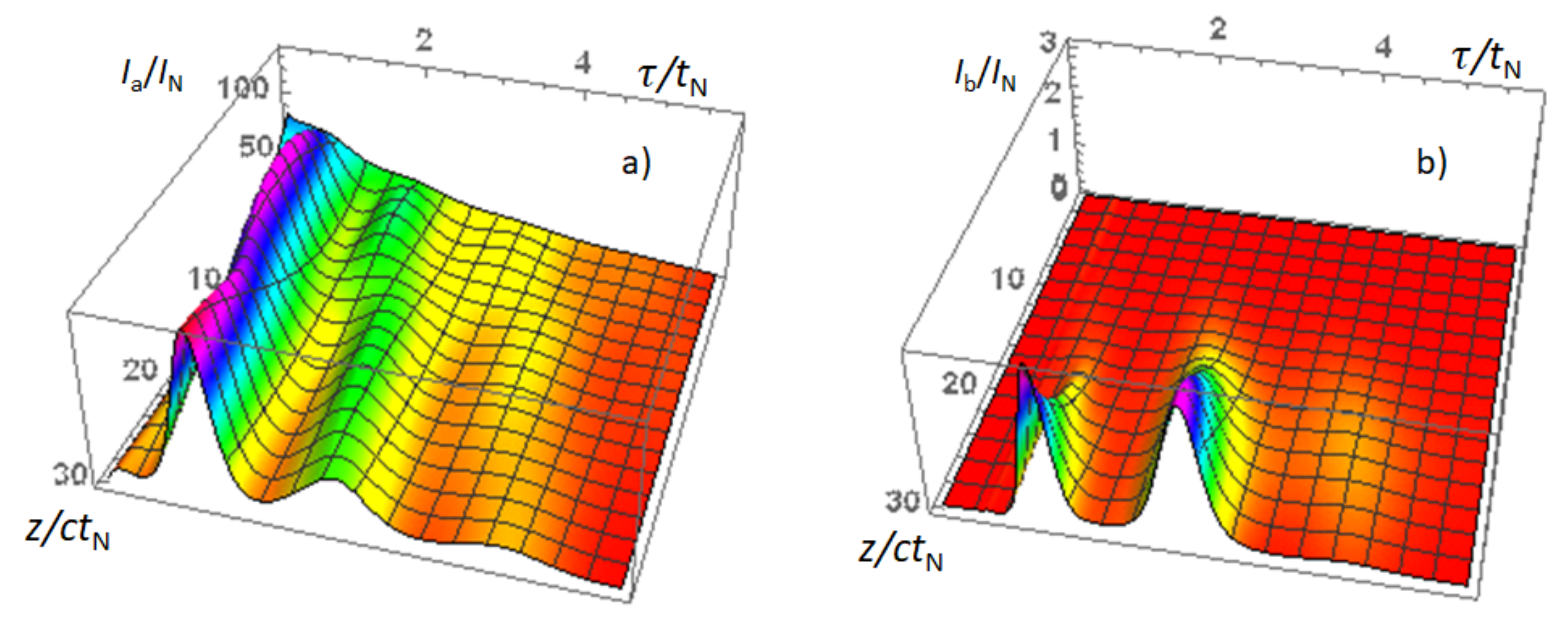} }
\caption{Spatio-temporal evolution of the intensity of the post-pump (a) and seed pulse (b). Parameters correspond to the pressure ratio $p/p_N=4$ and dimensionless pump amplitude $e_{ea0} = 5$. Other parameters are given in the text. }  \label{fig8}
\end{figure}

Analysis of the spatio-temporal evolution of the post-pump and seed pulses shown in Fig.~\ref{fig8} provides a better insight in the dynamics of soliton formation. First, the pump pulse is effectively absorbed in the plasma over distance of a few mm and forms three solitons, which propagate with a slower velocity. Seed amplification is correlated with the first two pump solitons, that is, the post-pump and the seed propagate together and form a joint soliton. The third soliton is too weak and the corresponding seed pulse is dissipated before reaching the plasma end.

The conditions of seed amplification are less favorable in the range of intermediate pressures, $p/p_N = 1-10$. This effect can be seen in Fig.~\ref{fig6}b as a slower increase of gain with  pressure (curve 1) or by a complete gain suppression (curve 2). In this pressure range the post-pump is already significantly absorbed and delayed in the plasma, but a soliton is not yet formed. The seed pulse is amplified as a single pulse with a delay but it not yet correlated to the pump post-pulse.

Therefore, seed amplification in the V-scheme occurs in two regimes of parametric amplification and soliton formation, both operating in a limited range of pressures. This range depends on the post-pulse amplitude, duration, frequency detuning and population on the level A. 

An increase of the pump amplitude extends the pressure range where seed gain increases monotonously with pressure up to the cutoff. At small pump amplitudes and large detunings, zones of parametric and soliton amplification are separated by an interval of pressures $p/p_N\simeq 1-10$ where gain either stagnates or suppressed. A similar behavior was reported in Ref.~\cite{Liu19}. Post-pulse duration has to be sufficiently long for enabling seed amplification. No amplification was observed numerically for pulse duration shorter than $2-3$~ps. Longer post-pulses favor stronger amplification and at longer seed delays. Seed amplification can be only observed at frequency detuning smaller than the corresponding pump Rabi frequency. This condition in dimensionless units reads: $e_{a0}|\Delta\omega_{ax}| \gtrsim 1$.

\section{Comparison with experimental results}\label{sec5}
The features of the V-scheme amplification demonstrated in Sec.~\ref{sec4} are in agreement with several experimental observations~\cite{Yao2013, Liu2013, Li2014, Liu15, Ivanov2020, Liu19}. An amplified signal at 391 nm that is delayed from the pump pulse by several ps has been observed in Ref.~\cite{Li2014, Liu15}. It displays an increase that is first growing approximately quadratic at low pressures until it reaches a maximum around $30-50$~mbar, and decreases at higher pressures~\cite{Liu15, Liu19}. Similar behavior can be seen in Fig.~\ref{fig6}b, curve 2. The measured in these papers gain is on the order of 100, which is also in agreement with Fig.~\ref{fig6}b. 

Simulations show that the temporal shape of the amplified signal depends sensitively on several parameters such a gas pressure, the length of the gain medium, amplitude, duration of the post pulse and the detuning of its frequency with respect to resonances. Because of a lack of knowledge on these parameters it is difficult make a quantitative comparison with experimental results. Nevertheless, the calculated temporal pulse shapes show striking similarity with several published results. For instance, the multiple temporal oscillations of the signal reported in Ref.~\cite{Li2014} compare well with signal pulse shapes shown in Figs.~\ref{fig4} and \ref{fig7}. Also, it is shown in Ref.~\cite{Liu19} that there is good agreement between calculated and measured temporal shapes at different pressures.

An amplification was also observed experimentally without injection of a seed pulse at 391 nm~\cite{Liu2013, Liu2017}. This is consistent with the V-scheme, which predicts amplification even if the initial coherent polarization B-X is on the order of $10^{-3}$. In this case, the B-X polarization is attributed to electron recollisions during the pump laser pulse duration~\cite{Liu2017}.

Signal amplification has been recently reported in air at normal pressure~\cite{Ivanov2020}. The pump pulse wavelength was at 950~nm and the lasing signal occurred at 428~nm. An amplified signal was reported that was delayed by 5.6~ps. According to our model, these conditions correspond to a soliton regime in a B(0)-X(1)-A(2) V-scheme arrangement. The corresponding pulse shapes are shown in Fig.~\ref{fig7}f. 

Finally, we note that it is also possible to obtain amplification with population inversion between the B and ground X(0) and X(1) states. This, however, requires higher intensities, above $\sim4\times10^{14}$\,W/cm$^{2}$~\cite{Li2019, ando2019}. These conditions can be obtained under tight focusing in thin gas jets~\cite{Britton2018}. 

\section{Conclusions}\label{sec6}
Our theoretical analysis shows that ionization of nitrogen gas by an intense femtosecond laser pulse at 800 nm is accompanied with a transfer of ionized molecules to higher excited levels due to resonance polarization coupling. This process depends on both the laser intensity and pulse duration. At moderate laser intensities (in the range  of $1-3\times10^{14}$\, W/cm$^{2}$), an ionization level of a few percents is reached, and one obtains an inversion between electronic levels B and A but no population inversion between B and X. In this case, a seed amplification can occur due to the coupling between A and B levels in a V-scheme. Two additional conditions must be fulfilled in order for the gain to take place (i) the main laser pulse has to be followed by a post-pulse of a few ps duration and an intensity 4 - 5 orders of magnitude smaller than the main pulse; and (ii) the spectrum of the post-pulse has to contain a component sufficiently close to one of the A-X transitions. The large number of rotational and vibrational levels in the excited ion facilitates this resonance condition. A three level V-scheme is sufficient for a description of this process as only one A-X transition closest to the spectral component of the laser post-pulse effectively participates in the coupling.

The seed amplification can be realized in two qualitatively different regimes: three-level parametric coupling or joint soliton propagation. In the former regime that occurs at pressures of less than 100 mbar, the post-pulse needs to be present during the whole process. This regime is experimentally observed in good agreement with our theory. The soliton regime is realized at higher pressures, where the amplified seed comes out synchronously with the post-pump in the form of a narrow pulse with a delay increasing with pressure. The soliton regime might have been observed in air at normal pressure.

Finally, at ionization level in excess of 40\% our theory shows that it is possible to obtain population inversion between the B and ground X state and to achieve direct amplification of a seed pulse at the B-X transition.

\begin{acknowledgments}
This research was partially supported by the Czech Republic MSMT targeted support of Large Infrastructures, ELI Beamlines Project LQ1606 of the National Programme of Sustainability II.
The authors acknowledge support from the project HiFI (CZ.02.1.01/0.0/0.0/15\_003/0000449), \linebreak ELITAS (CZ.02.1.01/0.0/0.0/16\_013/0001793) and ADONIS  (CZ.02.1.01/0.0/0.0/16\_019/0000789) from the European Regional Development Fund. The work is partially supported by the National Natural Science Foundation of China (Grants No. 11574213, 11904332), Innovation Program of Shanghai Municipal Education Commission (Grant No. 2017-01-07-00-07-E00007), and Shanghai Municipal Science and Technology Commission (Grant No. 17060502500).
\end{acknowledgments} 

\appendix
\section{Equations describing ionization and excitation of nitrogen molecules}\label{app1}
The temporal evolution of this 5 level system in a given laser electric field is described by the system of Bloch equations for density matrix involving 5 diagonal elements, $n_{a2}$, $n_{a3}$, $n_{b0}$, $n_{x0}$, $n_{x1}$, representing the population of correspondent states A(2), A(3), B(0), X(0) and X(1), and 6 off-diagonal elements, $d_{a2x0}$, $d_{a3x1}$, $d_{b0x0}$, $d_{b0x1}$, $d_{b0a2}$, and $d_{b0a3}$, accounting for polarization couplings between A and B states as shown in Fig.\ref{fig1}. This is a standard system for the polarization matrix~\cite{Lambopoulos2007} extended to five coupled states. 

The laser field responsible for molecule ionization and excitation is given by expression: 
$E_{\rm las} (t)=E_0 \cos(\omega_0 t) \sin(\pi t/t_{\rm las})$ where $E_0$ is the laser amplitude, $\omega_0$ is the carrier frequency and $t_{\rm las}$ is the pulse duration. The spatial dependence of the laser field is not considered in this model as the laser pulse is sufficiently strong and weakly modified in the gas. The five equations for the populations read:
\begin{eqnarray}
&& \partial_t n_{x0}=w_{ix0} (1-n_i)- 2 \mu_{ax}\hbar^{-1} E_{\rm las}  \sin\theta_n\, {\rm Im}d_{a2x0}- 2 \mu_{bx}\hbar^{-1}E_{\rm las} \cos\theta_n\,{\rm Im}d_{b0x0},  \label{eqa11} \\
&& \partial_t n_{x1}=w_{ix1} (1-n_i)- 2 \mu_{ax}\hbar^{-1} E_{\rm las}  \sin\theta_n\, {\rm Im}d_{a3x1}- 2 \mu_{bx}\hbar^{-1}E_{\rm las} \cos\theta_n\,{\rm Im}d_{b0x1},  \label{eqa12} \\
&&  \partial_t n_{a2}=w_{ia2} (1-n_i)+ 2 \mu_{ax} \hbar^{-1} E_{\rm las} \sin\theta_n\, {\rm Im}d_{a2x0},  \label{eqa21} \\
&&  \partial_t n_{a3}=w_{ia3} (1-n_i)+ 2 \mu_{ax} \hbar^{-1} E_{\rm las} \sin\theta_n\, {\rm Im}d_{a3x1},  \label{eqa22} \\
&& \partial_t n_{b0}=w_{ib0} (1-n_i)+ 2 \mu_{bx}\hbar^{-1} E_{\rm las} \cos\theta_n \,{\rm Im}d_{b0x0}+ 2 \mu_{bx}\hbar^{-1} E_{\rm las} \cos\theta_n \,{\rm Im}d_{b0x1}.  \label{eqa3}
\end{eqnarray}
They are completed with six equations for the corresponding polarizations:
\begin{eqnarray}
&& \partial_t d_{a2x0}=-i\omega_{a2x0} d_{a2x0}- i \mu_{ax}\hbar^{-1} E_{\rm las} \sin\theta_n\, (n_{a2}-n_{x0}) - i \mu_{bx} \hbar^{-1} E_{\rm las} \cos\theta_n \,d_{b0a2}^\star,  \label{eqa41} \\
&& \partial_t d_{a3x1}=-i\omega_{a3x1} d_{a3x1}- i \mu_{ax}\hbar^{-1} E_{\rm las} \sin\theta_n\, (n_{a3}-n_{x1}) - i \mu_{bx} \hbar^{-1} E_{\rm las} \cos\theta_n \,d_{b0a3}^\star,  \label{eqa42} \\
&& \partial_t d_{b0x0}=-i\omega_{b0x0} d_{b0x0}- i \mu_{bx}\hbar^{-1} E_{\rm las} \cos\theta_n\, (n_{b0}-n_{x0}) - i \mu_{ax} \hbar^{-1} E_{\rm las} \sin\theta_n \, d_{b0a2},  \label{eqa51} \\
&& \partial_t d_{b0x1}=-i\omega_{b0x1} d_{b0x1}- i \mu_{bx}\hbar^{-1} E_{\rm las} \cos\theta_n\, (n_{b0}-n_{x1}) - i \mu_{ax} \hbar^{-1} E_{\rm las} \sin\theta_n \, d_{b0a3},  \label{eqa52} \\
&& \partial_t d_{b0a2}=-i\omega_{b0a2} d_{b0a2}+ i\mu_{bx} \hbar^{-1} E_{\rm las} \cos\theta_n \,d_{a2x0}^\star - i \mu_{ax} \hbar^{-1} E_{\rm las} \sin\theta_n \,d_{b0x0},  \label{eqa61} \\
&& \partial_t d_{b0a3}=-i\omega_{b0a3} d_{b0a3}+ i\mu_{bx} \hbar^{-1} E_{\rm las} \cos\theta_n \,d_{a3x1}^\star - i \mu_{ax} \hbar^{-1} E_{\rm las} \sin\theta_n \,d_{b0x1}.  \label{eqa62}
\end{eqnarray}
Here $\hbar$ is the Planck constant, $w_{ia,b,x}$ are the ionization probabilities of neutral nitrogen molecule to the corresponding state and $n_i=n_{a2}+n_{a3}+n_{b0}+n_{x0}+ n_{x1}$ is the total ion density. The matrix elements are normalized to the initial density of neutral molecules. The damping rates are in the range of a few inverse picoseconds, they have no importance for the considered processes and have been neglected.

\section{Simple analytical solutions of a three-level system}\label{app2}
\subsubsection{Dark state}\label{sec331}
The system of equations~\eqref{eq11} -- \eqref{eq132} has a particular solution corresponding to constant electromagnetic field amplitudes and populations. This solution corresponds to zero polarizations, $p_{ax}=p_{bx}=0$. Then populations at the levels A, B and X are related to the field amplitudes by equations \eqref{eq11} and \eqref{eq12}:
$$ (n_a-n_x)\,e_a\sin\theta_i=-\frac12 p_{ba}^* e_b\cos\theta_i ,\qquad {\rm and}\qquad (n_b-n_x)\,e_b\cos\theta_i =-\frac12 p_{ba} e_a\sin\theta_i.$$
These relations imply that
$$ \frac{n_b-n_x}{n_a-n_x}=\frac{|e_a|^2\sin^2\theta_i}{|e_b|^2\cos^2\theta_i}=\tan^2\psi.$$ 
Taking into account that $n_a+n_b+n_x=1$, we find that such a solution may exist for arbitrary field amplitudes: 
\begin{equation} \label{eq10d}
n_a=(1-2n_x)\cos^2\psi+ n_x\sin^2\psi, \qquad  {\rm and}\qquad n_b=(1-2n_x)\sin^2\psi+ n_x\cos^2\psi.
\end{equation}
This solution implies that the population at the level X is sufficiently low, $n_x<0.5$. It corresponds to a ``dark state'' of a three level system, which allows propagation of both electromagnetic waves without absorption. That is, by injecting simultaneously the fields $e_a$ and $e_b$ one may maintain the population inversion in the states A and B for a long time, assuming that all ions are aligned at the same angle $\theta_i$ with respect to the electric field.

\subsubsection{Amplified spontaneous emission}\label{sec332}
Another known solution corresponds to the exponential amplification of a weak signal in a two level system in the conditions where there exists either a population inversion or a strong polarization. For example, Eqs.~\eqref{eq12} and \eqref{eq131} for an isolated B-X transition read:
$$ \partial_z e_b=\frac{r}2  p_{bx}\cos\theta_i, \qquad   \partial_\tau p_{bx}= (n_{b}-n_{x})\,e_b\cos\theta_i, \qquad \partial_\tau n_b= -\frac12 {\rm Re}(p_{bx}^*e_b)\,\cos\theta_i. $$
Neglecting population variation, $n_b=n_{b0}\approx {\rm const}$, a pair of equations for the electric field and polarization admits either an exponentially growing solution 
$$e_b \propto p_{bx0} \exp(\sqrt{2r(n_{b0}-n_{x0})z\tau}\,|\cos\theta_i|),$$ 
if $n_{b0}>n_{x0}$, or an oscillating solution if $n_{b0}<n_{x0}$. This growing solution may be realized at high pump intensities where a population inversion B-X is created by the pump pulse. 

\section{Angle-averaged equations for the V-scheme}\label{app3}
Equations~\eqref{eq11} -- \eqref{eq132} are averaged over the ion orientation angle $\theta_i$ assuming that the probability distribution $\mathcal{P}(\theta_i)$ is a time independent function with average value
$$ \langle\cos^2\theta_i\rangle = \frac12 \int_0^\pi \cos^2\theta_i\,\mathcal{P}(\theta_i)\, \sin\theta_i\,d\theta_i.$$ 
Following Refs.~\cite{Miyazaki2005, Zeng2009} we present in this paper simulation results with the value $ \langle\cos^2\theta_i\rangle =0.33$, but similar results have been obtained with other propability distributions.
By introducing the angle-averaged values for populations, $\bar{n}_{a,b,x}=\langle n_{a,b,x} \rangle$ and polarizations, $\bar{p}_{ax}=\langle p_{ax} \sin\theta_i\rangle$, $\bar{p}_{bx}=\langle p_{bx} \cos\theta_i\rangle$ and $\bar{p}_{ba}=\langle p_{ba} \sin\theta_i \cos\theta_i\rangle$, these equations can be written as follows:
\begin{eqnarray}
&&\partial_z e_a=\frac12 \bar{p}_{ax}, \qquad    \partial_\tau \bar{p}_{ax}=-\hat\gamma_{ax} \bar{p}_{ax}+ \langle(n_a-n_x) \sin^2\theta_i\rangle\, e_a +\frac{1}{2}  \bar{p}_{ba}^{\,*} e_b, \label{eq21}\\
&&\partial_z e_b=\frac{r}2  \bar{p}_{bx}, \qquad   \partial_\tau \bar{p}_{bx}=-\hat\gamma_{bx} \bar{p}_{bx}+ \langle(n_b-n_x)  \cos^2\theta_i\rangle\, e_b +\frac{1}{2}  \bar{p}_{ba} e_a,\label{eq22}\\
&& \partial_\tau \bar{n}_a= -\frac12 {\rm Re}(\bar{p}_{ax}^{\,*}e_a), \qquad  \partial_\tau \bar{n}_b= -\frac12{\rm Re}(\bar{p}_{bx}^{\,*}e_b), \label{eq231}\\
&& \partial_\tau \bar{p}_{ba}=-\hat\gamma_{ba} \bar{p}_{ba} -\frac{1}{2} \langle p_{ax}^*\cos^2\theta_i \sin\theta_i\rangle\, e_b-\frac{1}{2} \langle p_{bx} \sin^2\theta_i \cos\theta_i\rangle\, e_a^*.\label{eq232}
\end{eqnarray}
The higher order correlations in the right hand side of Eqs.~\eqref{eq21}, \eqref{eq22} and \eqref{eq232} are reduced in the lowest order. For example, $\langle n_a \sin^2\theta_i\rangle \approx \bar{n}_a \langle\sin^2\theta_i\rangle$ and $\langle p_{bx} \sin^2\theta_i \cos\theta_i\rangle \approx \bar{p}_{bx} \langle\sin^2\theta_i\rangle$.  Then, the system of equations takes a closed form:
\begin{eqnarray}
&&\partial_z e_a=\frac12 \bar{p}_{ax}, \qquad    \partial_\tau  \bar{p}_{ax}=-\hat\gamma_{ax} \bar{p}_{ax}+ (\bar{n}_a-\bar{n}_x)\, e_a  \langle\sin^2\theta_i\rangle +\frac{1}{2}  \bar{p}_{ba}^{\,*} e_b, \label{eq31}\\
&&\partial_z e_b=\frac{r}2  \bar{p}_{bx}, \qquad   \partial_\tau \bar{p}_{bx}=-\hat\gamma_{bx} \bar{p}_{bx}+ (\bar{n}_b-\bar{n}_x)\, e_b \langle\cos^2\theta_i\rangle +\frac{1}{2}  \bar{p}_{ba} e_a,\label{eq32}\\
&& \partial_\tau \bar{n}_a= -\frac12 {\rm Re}(\bar{p}_{ax}^{\,*}e_a), \qquad  \partial_\tau \bar{n}_b= -\frac12{\rm Re}(\bar{p}_{bx}^{\,*}e_b),\label{eq331} \\ && \partial_\tau \bar{p}_{ba}=-\hat\gamma_{ba} \bar{p}_{ba} -\frac{1}{2}\bar{p}_{ax}^{\,*}e_b \langle\cos^2\theta_i\rangle -\frac{1}{2} \bar{p}_{bx} e_a^* \langle\cos^2\theta_i\rangle.\label{eq332}
\end{eqnarray}
Solutions to this system of equations are presented in Sec.~\ref{sec4} for the case $\langle\cos^2\theta_i\rangle=0.33$.

\end{document}